\magnification=1250
\overfullrule=0pt
\baselineskip=19pt

\centerline{\bf Density-matrix renormalization-group studies of the spin-half} 
\centerline{\bf Heisenberg system with dimerization and frustration}
\vskip 0.2in

\centerline{R. Chitra$^\ast $, Swapan Pati$^{\ast \ast}$, H. R. 
Krishnamurthy$^{\ast , +} $,} 
\centerline{Diptiman Sen$^{\dagger, +}$ and S. Ramasesha$^{\ast 
\ast , +}$} 

\centerline{\it $^\ast $ Physics Department, Indian Institute of 
Science,}
\centerline{\it Bangalore 560012, India}

\centerline{\it $^{\ast \ast}$ Solid State and Structural Chemistry 
Unit,}
\centerline{\it Indian Institute of Science, Bangalore 560012, India}

\centerline{\it $^\dagger $ Centre for Theoretical Studies, Indian 
Institute of Science,}
\centerline{\it Bangalore 560012, India}

\centerline{\it $^{+}$ Jawaharlal Nehru Centre for Advanced Scientific 
Research,}
\centerline{\it Indian Institute of Science, Bangalore 560012,India} 

\vskip 0.3in
\line{\bf Abstract \hfill}

Using the density matrix renormalization group technique, we study the
ground state phase diagram and other low-energy properties of an 
isotropic antiferromagnetic spin-half chain with both dimerization and 
frustration, i.e., an alternation $\delta$ of the nearest neighbor 
exchanges and a next-nearest-neighbor exchange $J_2~$. For $\delta = 0$, 
the system is gapless for $J_2 < J_{2c} ~$ and has a gap for $J_2 > 
J_{2c} ~$ where $J_{2c}~$ is about $0.241$. For $J_2 = J_{2c}$, the gap 
above the ground state grows as $\delta$ to the power $0.667 \pm 0.001$. 
In the $J_2 - \delta$ plane, there is a disorder line $2 J_2 + \delta 
= 1$. To the left of this line, the peak in the static structure factor 
$S(q)$ is at $q_{max} = \pi$ (Neel phase), while to the right of the line, 
$q_{max}$ decreases from $\pi$ to $\pi / 2$ as $J_2$ is increased to large 
values (spiral phase). For $\delta = 1$, the system is equivalent to two 
coupled chains as on a ladder and it is gapped for all values of the 
interchain coupling.

\vskip 0.1in
\line{PACS numbers: 75.10.Jm, 75.50.Ee \hfill} 

\vfill
\eject

\line{\bf I. INTRODUCTION \hfill}
\vskip 0.2in

One of the most widely studied spin models is the isotropic spin-half 
Heisenberg antiferromagnetic chain. This model is Bethe ansatz
soluble$^{1,2}$ for uniform nearest-neighbor (NN) exchange constants.  
Since we can map the spin-half model in one dimension to a system of 
interacting spinless fermions by the Jordan-Wigner transformation, we
expect a Peierls type of dimerization in this system when it is coupled 
to phonons. The dimerized model, with exchange constants alternating as
$1 \pm \delta$, is not exactly solvable, although many numerical and
approximate analytic results exist for this model. There has also been
considerable interest in the spin-half chain$^3 $ with both NN and 
next-nearest-neighbor (NNN) antiferromagnetic interactions $J_2$. 
Interestingly, the ground state of this model can be solved exactly$^4 $ 
for $J_2 = 0.5$. The ground state is doubly degenerate for the ring and 
the two states are representable as the two Kekule structures of the 
valence-bond theory. 

The heuristic phase diagram (Fig. 1) of the ground state for a model that
incorporates the effects of both dimerization and frustration, i.e., 
the $J_2 - \delta$ model is known, although there has been no systematic 
study of this model in the $J_2 - \delta$ plane. The ground state is 
exactly known for points on the line $2 J_2 + \delta = 1$ $^5$. 
Analytical and numerical studies$^{6-8}$ of the model on the $\delta$=0 
line show a transition from a gapless phase for $J_2 < J_{2c} ~$ to a 
gapped phase for $J_2 > J_{2c} ~$. The value of $J_{2c} ~$ has been 
accurately computed to be $0.2411 \pm 0.0001$ in Ref. 8. The $J_2 =0, 
\delta > 0$ line corresponds to a dimerized spin chain. Another 
interesting line on the phase diagram is $\delta =1$ which corresponds 
(cf. Fig. 2) to coupled spin chains with $J_2 ~$ as the intrachain and 
$2$ as the interchain coupling constants. Little else is known about 
the general $J_2 -\delta$ model. Numerical studies of this model have 
been largely confined to exact low-lying states of small systems $N 
\leq 22$ on the $\delta=0$ line.

In this paper we present a systematic study of the various regions of this
phase diagram using the density matrix renormalization group (DMRG) 
method$^9$. Using this technique we have studied both open chains and rings 
of sizes ranging between $200$ and $300$ sites, depending upon the 
parameters of the model. In section II, we give a brief introduction to 
the method and present some computational details. In section III, we 
present our results and discuss them. 

\vskip 0.2in
\line{\bf II. THE HAMILTONIAN AND THE DMRG TECHNIQUE \hfill}
\vskip 0.2in

The Hamiltonian for the spin-half chain we have studied is given by
$$\eqalign{H ~&=~ H_0 ~+~ H_1 ~+~ H_2 \cr
H_0 ~&=~ \sum_{i=1}^{2N-1} ~ {\vec S}_i \cdot {\vec S}_{i+1} \cr
H_1 ~&=~ -~\delta ~\sum_{i=1}^{2N-1} ~(-1)^i {\vec S}_i \cdot 
{\vec S}_{i+1} \cr
H_2 ~&=~ J_2 ~\sum_{i=1}^{2N-2} ~{\vec S}_i \cdot {\vec S}_{i+2} ~, \cr}
\eqno(1)$$
where $J_2 ~$ is the NNN exchange and $\delta$ is the NN exchange alternation 
parameter. (Note that we have set the average NN exchange $J_1 = 1$). In 
the open chain with an even number of sites, the dimerization 
is chosen to make the exchange constant for the terminal bonds strong 
$(1+\delta)$. The topology of the chain being studied is shown in Fig. 2. 

The DMRG technique involves systematically building up the chain to a 
desired number of sites starting from a very short chain by adding two 
sites at a time. The initial chain of $2n$ sites, with $n$ a small enough 
integer, is diagonalized exactly and the reduced density matrix for the 
left $n$ sites is computed from the ground state of the $2n$ chain 
Hamiltonian by integrating over the states of the right $n$ sites. The 
density matrix is diagonalized and a matrix representation of the 
$n$-site Hamiltonian is obtained in a truncated basis using $m$ basis 
vectors which are the eigenvectors of the density matrix corresponding to
its $m$ largest eigenvalues. The Hamiltonian matrix for the $2n+2$ chain 
is then obtained in the $(2s+1)^2 m^2 ~$ dimensional direct product 
subspace obtained using the truncated basis of the left and the right 
parts of the $2n$ chain and the full space of the two additional spins
which are inserted in the middle. After obtaining the ground state of the 
$2n+2$ chain in the truncated basis, the density matrix of half the chain, 
now with $n+1$ sites, is obtained. The procedure is repeated to obtain the
ground state of the $2n+4$ chain. The iteration is stopped on reaching 
the desired chain length.

The DMRG technique works best for spin-half chains with short-range 
interactions. The accuracy depends crucially on the number of eigenvalues
of the density matrix, $m$, that are retained. For spin-half chains with
NN interactions only, $m=32$ gives results that approximate very well to 
the infinite system results. In systems with longer range interactions, 
it becomes necessary to keep track of the spin matrices in the most recent 
basis, corresponding to all the sites which interact with the new spins 
that are introduced at a given DMRG step. While the DMRG algorithm is 
devised to minimize the errors due to truncation of the Hamiltonian 
matrix, the algorithm does not necessarily retain the spin matrices very 
accurately. The limited success of the DMRG technique for two-dimensional
lattices can be attributed to this fact, since a two-dimensional lattice is 
topologically equivalent to a one-dimensional lattice with very long-range
interactions.  In the case of spin-half chains with NN and NNN interactions,
the spin matrices that appear in the Hamiltonian at any stage would have
undergone at most two transformations, and the results we find for m=64
compare well with exact numerical diagonalizations of chains upto 22 sites.

\vskip 0.2in
\line{\bf III. A FIELD THEORY FOR THE NEEL PHASE \hfill}
\vskip 0.2in

Before presenting our numerical results, we briefly discuss a field
theory for the low-energy and long-wavelength modes of the spin chain. 
In the Neel phase, the low-energy physics is well described by an $O(3)$ 
nonlinear sigma model (NLSM) with a topological term$^{10,12}$. (The 
field variable is an unit vector). This led Haldane to predict that 
integer and half-integer spin chains will have qualitatively different 
behaviour for $\delta = 0$, in that the former have a gap in their 
excitation spectrum while the latter do not. This prediction has received 
support from numerical calculations and experiments.

Since our primary interest in this paper lies in the spin-half chain, 
it is convenient to use a different approach which is specific to these 
chains. This is the technique of bosonization discussed by Luther and 
Peschel$^{11}$ and Affleck$^{12}$. In this approach, we first carry out a 
Jordan-Wigner transformation on the spin chain to obtain a model of 
interacting spinless fermions on a lattice
$$\eqalign{S_{i}^{z} ~=~ &\psi_{i}^{\dagger} \psi_{i} ~-~ {1 \over 2} \cr 
S_{i}^{+} ~=~ &\psi_{i} ~\exp ~[~i \pi ~\sum_{j < i} ~\psi_{j}^{\dagger}
\psi_{j} ~] ~. \cr} 
\eqno(2)$$
The spin-up and spin-down states at a site correspond respectively to states 
with and without a fermion. The continuum limit of the transformed 
Hamiltonian can be bosonized to obtain a tractable bosonic theory in $1+1$ 
dimensions. For our system, this just produces the theory of a free 
`relativistic' massless boson. The corresponding Lagrangian density is 
given by 
$${\cal L} ~=~ {1 \over 2} ~(\partial_{\mu} \phi )^2 ~,
\eqno(3)$$
where $\phi$ is a bosonic field. The scaling dimension of an operator 
$\exp (i \alpha \phi)$ is therefore $\alpha^2 / 4 \pi$. The relation 
between a spin operator and the continuum field $\phi$ is given by 
$$S^z (x) ~\sim ~(-1)^{x /a} ~\sin ({\sqrt {2 \pi}} ~\phi ) ~, 
\eqno(4)$$
where $a$ is the underlying lattice constant. Since the spin Hamiltonian 
is isotropic, it is sufficient to consider the $S^{z} S^{z}$ correlations. 
A straightforward evaluation of this correlation function yields
$$\langle ~S^z (x) ~S^z (0) ~\rangle ~\sim ~ {{(-1)^{x / a}} \over {\vert 
x \vert}} ~. 
\eqno(5)$$
This power-law correlation implies the absence of a gap above the ground 
state. Eq. (5) generally has corrections like $\log (x/a)$ because the 
field theory has a marginal operator ${\cal O}$. In the spin chain, this 
corresponds to the NNN exchange $J_2 ~$. 

There is a crtical value $J_{2c} ~$ such that ${\cal O}$ is marginally 
irrelevant for $J_2 < J_{2c} ~$ and marginally relevant for $J_2 > 
J_{2c} ~$. At $J_{2c} ~$, the field theory is exactly conformally 
invariant and there are no logarithmic corrections. The spectrum at that 
point is described by a specific conformal field theory. In particular, 
the gap between the ground state and the low-lying excited states scales 
with the chain length as $1/N$. Further, the first excited states with 
$S=0$ and $S=1$ are exactly degenerate. For $J_2 < J_{2c} ~$, the excited
$S=0$ state has a higher energy than the $S=1$ states. For $J_2 > 
J_{2c} ~$, the $S=0$ state becomes degenerate with the ground state as 
$N \rightarrow \infty$, and the triplet states are separated from the 
two ground states by a finite gap. One can therefore determine $J_{2c} ~$ 
accurately as the point where the singlet and triplet excited states 
are degenerate for large values of $N^{7,8}$. 

We can treat the spin-Peierls term $H_1 ~$ in Eq. (1) as a perturbation on 
$H_0 ~$ for small values of $\delta$. Using the above procedure, we obtain 
the bosonic representation of $H_1 ~$ in the continuum limit$^{12}$ as
$$H_1 ~\sim~ \delta ~\cos ({\sqrt {2 \pi}} ~\phi) ~.
\eqno(6)$$
This is a relevant operator with scaling dimension $1\over 2$ and it 
therefore produces a gap
$$\Delta ~\sim ~{\delta}^{2/3} ~.
\eqno(7)$$
Similarly, the change in the ground state energy caused by the perturbation
in (6) scales as
$$E_o (\delta) ~-~ E_o (0) ~\sim ~{\delta}^{4/3} ~.
\eqno(8)$$
These results are perturbative$^{14}$ and they are valid only for small 
$\delta$. Further, there are generally corrections of order $\log ~
(~\delta ~)$ in Eqs. (7) and (8) because of the presence of the marginal 
operator. However if we are exactly at the critical point $J_{2c} ~$, 
there are no logarithmic corrections to (7) and (8).

For $J_2 > J_{2c} ~$ and $\delta=0$, there is a gap between the two
degenerate ground states (both $S=0$) and the first excited state 
($S=1$). Near $J_{2c} ~$, the gap $\Delta$ has an {\it essential 
singularity} of the form
$$\Delta ~\sim ~ \exp ~(~- ~{A \over {J_2 - J_{2c}}} ~)
\eqno(9)$$
Hence the gap is numerically indistinguishable from zero unless $J_2 ~$
is greater than about $0.3$.

Before ending this section, we note that a different field theory is
required in the spiral phase (which, for $\delta = 0$, sets in beyond 
$J_2 = 0.5$ for spin-half). This field theory is a NLSM based on a 
$SO(3)$ matrix unlike the unit vector field in the NLSM for the Neel phase. 
The main feature of this new field theory is that there is no topological
term and therefore no qualitative difference between integer and half-integer
spin chains$^{13}$. There should be gap above the ground state(s) in 
either case.

\vskip 0.2in
\line{\bf IV. RESULTS AND DISCUSSION \hfill}
\vskip 0.2in

We now present the phase diagram of the model and the various results
obtained by us using DMRG. We also compare these with the existing 
analytical and numerical results. This is done in four subsections:
A. The frustrated model with $\delta=0$; B. The dimerized model with 
$J_2 ~=$ constant; C. The general $J_2 - \delta$ model; and, D. Coupled 
spin chains with $\delta =1$.

\vskip 0.2in
\line{\bf A. The frustrated spin chain $(J_2 > 0, \delta =0)$ \hfill}
\vskip 0.2in
 
Very few exact results are known for this model. The ground state 
is exactly solvable for $J_2 =0.5$. For cyclic boundary conditions, 
the ground state is doubly degenerate with the wave functions being given 
by the two possible Kekule structures, namely, the two patterns of singlets 
which can be formed by neighboring sites.
$$\eqalign{\psi_1 ~&=~ [1,2] ~[3,4]~.... ~[2N-1 ,2N] \cr
\psi_2 ~&=~ [2N,1] ~[2,3]~ ....~ [2N-2,2N-1] ~, \cr}
\eqno(10)$$
where $[i,j]$ denotes the normalized singlet combination of the spins on 
sites $i$ and $j$.
 
The field theory studies by Haldane on the frustrated model revealed the 
existence of a transition from a gapless phase for $J_2 < J_{2c} ~$ to a 
gapped phase for $J_2 > J_{2c} ~$. He estimated $J_{2c} ~$ to be $0.16$. 
A direct computation of the gap obtained by exact diagonalization of 
small systems$^3$ placed $J_{2c} ~$ near $0.30$. However this turns out
to be an unreliable method because of the essential singularity in Eq. (9) 
as $J_2 ~$ approaches $J_{2c} ~$ from above. The most reliable method of
computing $J_{2c} ~$ is the one based on the crossing of the excited singlet
and triplet states extrapolated to infinite system size through finite 
size scaling. Ref. 8 obtained a value of $J_{2c} ~=0.2411 \pm 0.0001$ 
by this method. In the next subsection, we use this value of $J_{2c} ~$
to study the relation between the gap and $\delta$.

We have confirmed that the ground state is doubly degenerate for $J_2 >
J_{2c} ~$, as required by the Lieb-Schultz-Mattis theorem for gapped 
spin-half systems. Fig. 3 summarizes our results for the extrapolated 
gap for $0 < J_2 \le 1$. The gap increases for $J_2 > J_{2c} ~$ as a 
function of $J_2 ~$ until about $J_2 =0.7$ beyond which it decreases. 
In the limit of very large $J_2 ~$, the gap should once again vanish 
since the system then consists of two decoupled nearest neighbor chains. 
We find that this behaviour sets in even at a relatively small value of 
$J_2 =0.8$. Fig. 4 shows the convergence of the gap with increasing 
chain length in the gapped ($1/N^2 ~$) as well as in the gapless ($1/N$)
regions, and illustrates the validity of our extrapolation procedures.

\vskip 0.2in
\line{\bf B. Dimerized model $(J_2 = {\rm constant}, \delta > 0)$ \hfill} 
\vskip 0.2in

At $J_{2c} ~$, we have calculated the gap $\Delta$ for various values of 
$\delta$ on spin chains of upto 300 sites with the open boundary condition. 
The ground state is a singlet and the lowest excitation is a triplet. The 
triplet is therefore obtained as the lowest energy state in the $S^z =1$ 
sector within the DMRG formalism. The minimum chain length at which the 
infinite chain behaviour can be expected to set in scales as the 
correlation length and hence depends inversely on $\Delta$. For 
$\delta=0.007$ (which is the smallest $\delta$ we have studied), the 
infinite chain behaviour is expected for lengths greater than about 
$200$ sites. 

Fig. 5 shows that the log - log plot of the gap versus $\delta$ is 
linear. The exponent for the gap is given by the slope of this 
straight line which is $0.667 \pm 0.001$. Previous numerical
studies using exact diagonalization of spin chains with upto 22 sites 
had placed the value of this exponent between $0.9$ and $1.0$ $^{15,16}$,
while those using finite size scaling placed it closer to $0.75^{17}$.
However these studies were all at $J_2 = 0$, hence they were subject to 
errors due to the logarithmic corrections. Fig. 6 shows the log - log
plot of the change in the ground state energy $E_o (\delta ) - E_o (0)~$
versus $\delta$. The slope of the straight line is $1.251 \pm 0.001$
which is somewhat different from the field theory result of $4/3$.

For a `relativistic' massive theory which differs from a massless theory
by a small perturbation (by small we mean that the `velocity of light' 
is not changed), the mass gap $\Delta$ and the correlation length $\xi$ 
should be inversely related to each other. In Fig. 7, we plot the product 
$\xi \Delta$ versus $\delta$ for $J_2 = 0$, and find that this does appear 
to be the case.  

\vskip 0.2in
\line{\bf C. The general $J_2 - \delta$ model \hfill}
\vskip 0.2in
 
The ground state of the model is known to be exactly solvable along the
line $2 J_2 + \delta =1$. The presence of the dimerization $\delta$ lifts 
the degeneracy of the two ground states at $J_2 =0.5$. One of the Kekule 
states continues to be the ground state while the other is no longer an 
eigenstate of the Hamiltonian. Choosing the state $\psi_1$ in Eq. (9) as 
a trial wave function, we obtain a variational energy per site equal to
$$E_0 ~=~ -{3 \over 8} ~(1 + \delta) ~.
\eqno(10)$$
Using the Rayleigh-Ritz variational principle, it is then easy to show 
that $E_0$ saturates the lower bound to $\langle ~H~ \rangle$ provided$^5$
$2J_2 + \delta =1$. Fig. 8 shows the variation of the gap along this line.  

The general model exhibits a gap in the excitation spectrum for all 
points $(J_2 ,\delta)$ not covered by the special cases discussed above.
The ground state energies for various values of $J_2 ~$ and $\delta$ are
given in Table 1.

To characterize the ground state further, we computed the
equal time spin-spin correlation function at many points in the phase
diagram. The system sizes chosen for these calculations were varied with 
$J_2$ and $\delta$ so as to reasonably approximate the infinite chain 
behaviour. We then computed the static structure factor $S(q)$ defined as 
the Fourier transform of the correlation function. Some information about the 
type of long-range order (LRO), if any, can be obtained from the dependence 
of $S(q)$ on $q$. The classical limit ($S \rightarrow \infty $) predicts 
Neel order for $J_2 <0.25$ and a (coplanar) spiral order for $J_2 >0.25$. 
In other words, $S(q)$ has a peak at $q_{max}=\pi$ for $J_2 <0.25$ and 
$q_{max}=\cos^{-1} (-1 / 4J_2 )$ for $J_2 > 0.25$. The periodicity of the 
ground state is $ 2\pi / q_{max} ~$.

The quantum model has no LRO but exhibits a short-range order characterized
by some $q_{max}$ (and a finite correlation length if there is a gap in the 
spectrum). Earlier numerical studies$^3$ of the $S=1/2$ system revealed 
marked deviations from the large-$S$ (classical) results. They showed that 
$q_{max}=\pi$ for $ J_2 \leq 0.5$ and $q_{max} < \pi$ for $J_2 >0.5$, with 
$q_{max}$ approaching $\pi / 2$ for large $J_2 ~$.

We have calculated $S(q)$ for rings of sizes $100$ and $150$ at the points
indicated in Fig. 9. The structure factor peaks at $q_{max}= \pi$ for 
all points along the $J_2 =0$ line. On the $\delta=0$ line, $S(q)$ peaks at 
$q_{max}=\pi$ for $J_2 \leq 0.5$ and $q_{max}< \pi$ for $J_2 >0.5$ with 
$q_{max}$ approaching $\pi / 2$ for very large $J_2 ~$ as shown in Fig. 10. 
In the full $J_2 -\delta$ plane, $S(q)$ shows a very interesting change 
of behaviour across the line $2 J_2 + \delta = 1$. For points to the left 
of the line, $S(q)$ peaks at $q_{max}=\pi$ while for points to the right 
of the line, $S(q)$ peaks at $q_{max} < \pi$. However, for points close to 
the line on the right hand side, $S(q)$ shows a very broad distribution 
near $q = \pi$ and we cannot identify a clear maximum in the plot (Fig. 
11).

The correlation length $\xi$ is a minimum along the entire line $2 J_2 + 
\delta =1$, reflecting a highly disordered ground state. The main features 
in the $S(q)$ plots seem to be in conformity with this behavior of $\xi$.
We can therefore call $2J_2 + \delta =1$ a disorder line. The disorder 
line separates the ground states with $q_{max}=\pi$ and $q_{max} <\pi$ 
as shown in Fig. 9.

\vskip 0.2in
\line{\bf D. Coupled spin chains ($\delta =1$) \hfill}
\vskip 0.2in

For $\delta=1.0$ and $J_2 > 0$, the model corresponds to two coupled spin 
chains as discussed earlier. The phase diagram of coupled spin-half chains 
was studied in Ref. 18. We have examined the dependence of the gap on the 
scaled strength of the interchain coupling $J_{i} ~\equiv ~2 / J_2$. This 
was calculated for a $2 \times 100$ ladder. In Fig. 12, we see a gap in 
the spectrum even for very small values of $J_i$ as shown. Our results 
for the excitation gaps are in agreement with the studies of Barnes et 
al$^{19}$ which suggested the presence of a non-zero energy gap between 
the singlet ground state and the triplet excited states for all 
antiferromagnetic interchain couplings.

The convergence of the ground state energy per site for this system is 
shown in Fig. 13. The fluctuation in the ground state energy with the total
number of sites is because of the alternation between even and odd
number of sites on a single chain. The amplitude of the fluctuation is 
large for weak interchain couplings and gets damped as the coupling
between the chains increases. The energy per site when the intrachain
and interchain exchange constants are equal is $0.5780$. This value is 
intermediate between the ground state energy per site for the isotropic 
Heisenberg anitferromagnet on the two-dimensional square lattice ($-0.67$) 
and the ground state energy per site on a chain ($-0.4431$). The advantage 
of our method is that it is still treated as a single chain problem and 
thus avoids the pit falls of DMRG in two dimensions. The accuracy of the 
method for the ladder is the same as that for the single chain. 

\vskip 0.2in
\line{\bf V. SUMMARY \hfill}
\vskip 0.2in

To conclude, we have studied the NN and NNN antiferromagnetic spin-half 
Heisenberg chain with a dimerization, $\delta$, in the NN exchange. When 
$\delta$ is zero, there is a critical value $J_{2c} ~= 0.2411$ below 
which the system is gapless and above which the system is gapped. 
For the dimerized chain with the critical value of the NNN interaction, 
the energy gap above the ground state scales as $\delta$ to the power 
$0.667 \pm 0.001$, while the change in the ground state energy $E_o 
(\delta) - E_o $ scales with the power $1.251 \pm 0.001$. In the $J_2 -
\delta$ plane, we find a disorder line $2J_2 + \delta =1$ to the left of 
which the static structure factor $S(q)$ peaks at $q_{max}=\pi$. To the 
right of this line, $q_{max}$ gradually decreases from $\pi$ to $\pi / 2$ 
for large $J_2 ~$. For $\delta=1$, the model corresponds to coupled chains 
as on a ladder. The system is gapped for all values of antiferromagnetic 
$J_i ~$. The energy per site is $-0.57804$ for $J_{i} =1.0$ which lies 
in between the values for the one-dimensional and two-dimensional (square 
lattice) NN antiferromagnetic systems. 

\vskip 0.8in
\line{\bf REFERENCES \hfill}
\vskip 0.2in

\noindent
\item{1.}{J. des Cloizeaux and J.J. Pearson, Phys. Rev. {\bf 128}, 2131
(1962) and references therein.}

\noindent
\item{2.}{J.C. Bonner and M.E. Fisher, Phys. Rev. {\bf 135}, A640 
(1964).}

\noindent
\item{3.}{T. Tonegawa and I. Harada,  J. Phys. Soc. Jpn. {\bf 56}, 2153 
(1987).}

\noindent
\item{4.}{C. K. Majumdar and D. K. Ghosh, J. Math. Phys. {\bf 10},
1388 (1969); C. K. Majumdar, J. Phys. C {\bf 3}, 911 (1970).}

\noindent
\item{5.}{B. S. Shastry and B. Sutherland, Phys. Rev. Lett. {\bf 47},
964 (1981).}

\noindent
\item{6.}{F. D. M. Haldane, Phys. Rev. B {\bf 25}, 4925 (1982);
R. Jullien and F. D. M. Haldane, Bull. Am. Phys. Soc. {\bf 28}, 344
(1983)}

\noindent
\item{7.}{I. Affleck, D. Gepner, H. J. Schulz and T. Ziman, J. Phys.
A {\bf 22}, 511 (1989).}

\noindent
\item{8.}{K. Okamoto and K. Nomura, Phys. Lett. A {\bf 169}, 433 
(1992).}

\noindent
\item{9.}{S. R. White,  Phys. Rev. Lett. {\bf 69}, 2863 (1992);
Phys. Rev. B {\bf 48}, 10345 (1993).}

\noindent
\item{10.}{F. D. M. Haldane, Phys. Lett. {\bf 93A}, 464 (1983); 
Phys. Rev. Lett. {\bf 50} 1153 (1983).}

\noindent
\item{11.}{A. Luther and I. Peschel, Phys. Rev. B {\bf 12}, 3908 (1975).}

\noindent
\item{12.}{I. Affleck, in {\it Fields, Strings and Critical Phenomena}, 
eds. E. Brezin and J. Zinn-Justin (North-Holland, Amsterdam, 1989); I. 
Affleck and F.D.M. Haldane, Phys. Rev. B {\bf 36}, 5291 (1897).}

\noindent
\item{13.}{S. Rao and D. Sen, Nucl. Phys. B {\bf 424}, 547 (1994); 
D. Allen and D. Senechal, to appear in Phys. Rev. B (1995).}

\noindent
\item{14.}{M. C. Cross and D. S. Fisher, Phys. Rev. B {\bf 19}, 402 
(1979).}

\noindent
\item{15.}{S. Inagaki and  H. Fukuyama, J. Phys. Soc. Jpn. {\bf 52}, 2504
(1983); S. Ramasesha and Z. G. Soos, Solid State Commun. {\bf 46}, 509
(1983).}

\noindent
\item{16.}{Z. G. Soos, S. Kuwajima, J.E. Mihalick, Phys. Rev. B {\bf 32}, 
3124 (1985).}

\noindent
\item{17.}{J. C. Bonner and H.W.J. Blote, Phys. Rev. B {\bf 25}, 6959
(1982).}
 
\noindent
\item{18.}{S. P. Strong and A. J. Millis, Phys. Rev. Lett. {\bf 69},
2419 (1992).}

\noindent 
\item{19.}{T. Barnes, E. Dagotto, J. Riera and E. S. Swanson, Phys. Rev. 
B {\bf 47}, 3196 (1993).}

\vfill
\eject

\line{\bf Table Captions \hfill}
\vskip 0.2in

\noindent
\item{1.}{Ground state energy per site $E_o ~$ for various values of
$J_2 ~$ and $\delta$.}

\noindent
\item{2.}{Ground state energy per site for coupled spin chains for
various values of $J_i ~$.}

\vfill
\eject

\vbox{\tabskip=0pt \offinterlineskip
\def\tablerule{\noalign{\hrule}}
\halign to350pt{\strut#& \vrule#\tabskip=1em plus2em&
\hfil#& \vrule#& \hfil#& \vrule#& \hfil#& \vrule#\tabskip=0pt \cr
\tablerule
&&\omit&&\omit&&\omit& \cr
&&\omit\hidewidth$J_2$\hidewidth&&\omit\hidewidth $\delta$\hidewidth&&
\omit\hidewidth $E_o$\hidewidth& \cr
&&\omit&&\omit&&\omit& \cr
\tablerule
&&\omit&&\omit&&\omit& \cr
&&\omit\hidewidth$0.2411$\hidewidth&&\omit\hidewidth $0.000$\hidewidth&&
\omit\hidewidth $-0.401866$\hidewidth& \cr
&&\omit\hidewidth$0.2411$\hidewidth&&\omit\hidewidth $0.007$\hidewidth&&
\omit\hidewidth $-0.402681$\hidewidth& \cr
&&\omit\hidewidth$0.2411$\hidewidth&&\omit\hidewidth $0.014$\hidewidth&&
\omit\hidewidth $-0.403766$\hidewidth& \cr
&&\omit\hidewidth$0.2411$\hidewidth&&\omit\hidewidth $0.020$\hidewidth&&
\omit\hidewidth $-0.404844$\hidewidth& \cr
&&\omit\hidewidth$0.2411$\hidewidth&&\omit\hidewidth $0.028$\hidewidth&&
\omit\hidewidth $-0.406430$\hidewidth& \cr
&&\omit\hidewidth$0.2411$\hidewidth&&\omit\hidewidth $0.040$\hidewidth&&
\omit\hidewidth $-0.409051$\hidewidth& \cr
&&\omit\hidewidth$0.2411$\hidewidth&&\omit\hidewidth $0.057$\hidewidth&&
\omit\hidewidth $-0.413132$\hidewidth& \cr
&&\omit\hidewidth$0.2411$\hidewidth&&\omit\hidewidth $0.080$\hidewidth&&
\omit\hidewidth $-0.419154$\hidewidth& \cr
&&\omit\hidewidth$0.2411$\hidewidth&&\omit\hidewidth $0.160$\hidewidth&&
\omit\hidewidth $-0.442862$\hidewidth& \cr
&&\omit\hidewidth$0.2411$\hidewidth&&\omit\hidewidth $0.230$\hidewidth&&
\omit\hidewidth $-0.465728$\hidewidth& \cr
&&\omit\hidewidth$0.2411$\hidewidth&&\omit\hidewidth $0.320$\hidewidth&&
\omit\hidewidth $-0.496844$\hidewidth& \cr
\tablerule
&&\omit&&\omit&&\omit& \cr
&&\omit\hidewidth$0.10$\hidewidth&&\omit\hidewidth $0.0$\hidewidth&&
\omit\hidewidth $-0.42517$\hidewidth& \cr
&&\omit\hidewidth$0.20$\hidewidth&&\omit\hidewidth $0.0$\hidewidth&&
\omit\hidewidth $-0.40885$\hidewidth& \cr
&&\omit\hidewidth$0.25$\hidewidth&&\omit\hidewidth $0.0$\hidewidth&&
\omit\hidewidth $-0.40045$\hidewidth& \cr
&&\omit\hidewidth$0.30$\hidewidth&&\omit\hidewidth $0.0$\hidewidth&&
\omit\hidewidth $-0.39284$\hidewidth& \cr
&&\omit\hidewidth$0.40$\hidewidth&&\omit\hidewidth $0.0$\hidewidth&&
\omit\hidewidth $-0.38028$\hidewidth& \cr
&&\omit\hidewidth$0.50$\hidewidth&&\omit\hidewidth $0.0$\hidewidth&&
\omit\hidewidth $-0.37500$\hidewidth& \cr
&&\omit\hidewidth$0.60$\hidewidth&&\omit\hidewidth $0.0$\hidewidth&&
\omit\hidewidth $-0.38079$\hidewidth& \cr
&&\omit\hidewidth$0.70$\hidewidth&&\omit\hidewidth $0.0$\hidewidth&&
\omit\hidewidth $-0.39711$\hidewidth& \cr
&&\omit\hidewidth$0.80$\hidewidth&&\omit\hidewidth $0.0$\hidewidth&&
\omit\hidewidth $-0.42138$\hidewidth& \cr
&&\omit\hidewidth$1.00$\hidewidth&&\omit\hidewidth $0.0$\hidewidth&&
\omit\hidewidth $-0.48565$\hidewidth& \cr
\tablerule
&&\omit&&\omit&&\omit& \cr
&&\omit\hidewidth$0.45$\hidewidth&&\omit\hidewidth $0.07$\hidewidth&&
\omit\hidewidth $-0.40130$\hidewidth& \cr
&&\omit\hidewidth$0.48$\hidewidth&&\omit\hidewidth $0.10$\hidewidth&&
\omit\hidewidth $-0.41281$\hidewidth& \cr
&&\omit\hidewidth$0.55$\hidewidth&&\omit\hidewidth $0.10$\hidewidth&&
\omit\hidewidth $-0.41610$\hidewidth& \cr
&&\omit\hidewidth$0.15$\hidewidth&&\omit\hidewidth $0.20$\hidewidth&&
\omit\hidewidth $-0.46329$\hidewidth& \cr
&&\omit\hidewidth$0.25$\hidewidth&&\omit\hidewidth $0.35$\hidewidth&&
\omit\hidewidth $-0.50727$\hidewidth& \cr
&&\omit\hidewidth$0.40$\hidewidth&&\omit\hidewidth $0.50$\hidewidth&&
\omit\hidewidth $-0.56611$\hidewidth& \cr
&&\omit\hidewidth$0.15$\hidewidth&&\omit\hidewidth $0.60$\hidewidth&&
\omit\hidewidth $-0.60033$\hidewidth& \cr
&&\omit\hidewidth$0.20$\hidewidth&&\omit\hidewidth $0.80$\hidewidth&&
\omit\hidewidth $-0.67613$\hidewidth& \cr
&&\omit\hidewidth$0.30$\hidewidth&&\omit\hidewidth $0.80$\hidewidth&&
\omit\hidewidth $-0.67966$\hidewidth& \cr
&&\omit\hidewidth$0.48$\hidewidth&&\omit\hidewidth $0.80$\hidewidth&&
\omit\hidewidth $-0.69256$\hidewidth& \cr
&&\omit&&\omit&&\omit& \cr
\tablerule}}

\vskip 0.2in
\centerline{\bf Table 1}

\vfill
\eject

\vbox{\tabskip=0pt \offinterlineskip
\def\tablerule{\noalign{\hrule}}
\halign to350pt{\strut#& \vrule#\tabskip=1em plus2em&
\hfil#& \vrule#& \hfil#& \vrule#\tabskip=0pt \cr
\tablerule
&&\omit&&\omit& \cr
&&\omit\hidewidth $J_i$\hidewidth&&\omit\hidewidth $E_o$
\hidewidth& \cr
&&\omit&&\omit& \cr
\tablerule
&&\omit&&\omit& \cr
&&\omit\hidewidth $0.020$\hidewidth&&\omit\hidewidth $-0.44320$
\hidewidth& \cr
&&\omit\hidewidth $0.040$\hidewidth&&\omit\hidewidth $-0.44337$
\hidewidth& \cr
&&\omit\hidewidth $0.066$\hidewidth&&\omit\hidewidth $-0.44434$
\hidewidth& \cr
&&\omit\hidewidth $0.133$\hidewidth&&\omit\hidewidth $-0.44320$
\hidewidth& \cr
&&\omit\hidewidth $0.200$\hidewidth&&\omit\hidewidth $-0.44741$
\hidewidth& \cr
&&\omit\hidewidth $1.000$\hidewidth&&\omit\hidewidth $-0.57804$
\hidewidth& \cr
&&\omit&&\omit& \cr
\tablerule}}

\vskip 0.2in
\centerline{\bf Table 2}

\vfill
\eject

\line{\bf Figure Captions \hfill}
\vskip 0.2in

\noindent
\item{1.}{Heuristic phase diagram for the spin-half chain. The solid line
along $\delta = 0$ from $J_2 = 0$ to $J_{2c} ~$ is gapless; the rest of the
diagram is gapped. The line $2J_2 + \delta = 1$ separates the Neel phase
from the spiral phase.}

\noindent
\item{2.}{Schematic picture of the antiferromagnetic exchanges in the 
chain.}

\noindent
\item{3.}{Dependence of the gap $\Delta$ on $J_2$ for $\delta = 0$.}

\noindent
\item{4.}{Convergence of the singlet-triplet gap $\Delta$ with $1 / N$ 
for $J_2 =0.5$ (gapped system) and $J_2 =0.2$ (gapless) at $\delta = 0$.}

\noindent
\item{5.}{Log-log plot of the gap $\Delta$ versus $\delta$ for $J_2 = 
0.2411$.}

\noindent
\item{6.}{Log-log plot of the change in the ground state energy $E_o
(\delta ) - E_o (0)$ versus $\delta$ for $J_2 = 0.2411$.}

\noindent
\item{7.}{Plot of the product of the correlation length $\xi$ and
the gap $\Delta$ versus $\delta$ for $J_2 = 0$.}

\noindent
\item{8.}{Gap $\Delta$ versus $\delta$ along the line $2 J_2 + \delta = 
1$.}

\noindent
\item{9.}{Behaviour of the structure factor $S(q)$ in the $J_2 - \delta$ 
plane.}

\noindent
\item{10.}{Plot of $q_{max}$ (in degrees) versus $J_2 ~$ for $\delta = 0$.}

\noindent
\item{11.}{$S(q)$ versus $q$ (in degrees) for various values of $J_2$ and 
$\delta$.}

\noindent
\item{12.}{Gap versus the interchain coupling $J_i ~$ for coupled chains.} 

\noindent
\item{13.}{Convergence of the ground state energy per spin $E_o ~$ with 
system size for coupled chains with $J_i =0.04$.}

\end